\documentclass[10pt,a4paper]{article}
\usepackage{amsmath}
\usepackage{amsfonts}
\usepackage{amssymb}
\usepackage{amsthm}
\usepackage[square]{natbib}
\usepackage{hyperref} % Remove hyperref to remove boxes
\usepackage[a4paper, total={6.5in, 8in}]{geometry}
\usepackage{color}
\usepackage{graphicx}
\usepackage{float}
\usepackage{doi}
\hypersetup{hidelinks}
\usepackage{tikz}
\usepackage[export]{adjustbox}
\usepackage{layouts}
\usepackage{psfrag}
\usepackage{stmaryrd}
\usepackage{comment}
\usepackage{booktabs}
\usetikzlibrary{shapes.geometric, arrows}

\newcommand{\nn}{\mathcal{NN}}
\newcommand{\hen}{\text{Hencky}}
\newcommand{\ssve}{\text{ssve}}
\newcommand{\nem}{\text{nematic}}

\newtheorem{theorem}{Theorem}

\newtheorem{lemma}[theorem]{Lemma}

\newtheorem{remark}{Remark}
\usepackage{mathtools}
\usepackage{multicol}
\usepackage{pgfplots}
\usetikzlibrary{arrows,shapes,positioning}
\usepgfplotslibrary{groupplots}
\usepackage{multirow}
\usepackage{caption}
\usepackage{subcaption}

\newcommand{\R}{\mathbb{R}}

\newcommand{\SO}{\mathcal{SO}}

\newcommand{\m}{m}
\usepackage[hang,flushmargin,symbol]{footmisc}

\begin{document}
\begin{center}
%\noindent\textbf{{\Large{Incorporating sufficient physical information into artificial neural networks: a guaranteed improvement via physics-based Rao-Blackwellization}}}
\textbf{\Large{Input convex neural networks: universal approximation theorem and implementation for isotropic polyconvex hyperelastic energies}}
~\\
~\\
~\\
Gian-Luca Geuken$^1$, Patrick Kurzeja$^1$, David Wiedemann$^2$, J\"orn Mosler$^{1,}$\footnote{\noindent Corresponding author\\Email adress: joern.mosler@tu-dortmund.de}
~\\
~\\
$^1$\textit{Institute of Mechanics, Department of Mechanical Engineering,
TU Dortmund University, Leonhard-Euler-Str. 5, 44227 Dortmund, Germany}\\
$^2$\textit{Applied Analysis, Faculty of Mathematics,
TU Dortmund University, Vogelpothsweg 87, 44227 Dortmund, Germany}
~\\
~\\
\end{center}
Abstract:
This paper presents a novel framework of neural networks for isotropic hyperelasticity that enforces necessary physical and mathematical constraints while simultaneously satisfying the universal approximation theorem. The two key ingredients are an input convex network architecture and a formulation in the elementary polynomials of the signed singular values of the deformation gradient. In line with previously published networks, it can rigorously capture frame-indifference and polyconvexity -- as well as further constraints like balance of angular momentum and growth conditions. However and in contrast to previous networks, a universal approximation theorem for the proposed approach is proven. To be more explicit, the proposed network can approximate any frame-indifferent, isotropic polyconvex energy (provided the network is large enough). This is possible by working with a sufficient and necessary criterion for frame-indifferent, isotropic polyconvex functions. Comparative studies with existing approaches identify the advantages of the proposed method, particularly in approximating non-polyconvex energies as well as computing polyconvex hulls.
~\\
~\\
Keywords: artificial neural networks, hyperelasticity, polyconvexity, material modeling, singular values
\thispagestyle{empty}
\clearpage

%%%%%%%%%%%%%%%%%%%%%%%%%%%%%%%%%%%%%%%%%%%
\section{Introduction}
%%%%%%%%%%%%%%%%%%%%%%%%%%%%%%%%%%%%%%%%%%%

%%%%%%%%%%%%%%%%%%%%%%%%%%%%%%%%%%%%%%%%%%%
\subsection{Motivation for physics-based neural networks in constitutive modeling}

Recent developments have shown that data-driven and machine learning methods such as neural networks are indeed useful alternatives to classical formulations of physical models \citep{kirchdoerfer2016,linden2023,bartel2023}. They pave the way for automatic model generation and offer enormous flexibility in terms of functional relationships \citep{thakolkaran2022}. A potential for the improvement of such approaches was found in embedding physical knowledge and principles into them \citep{kumar2022,peng2021,moseley2022,geuken2024}.\\
The present work specifically aims at the incorporation of physical principles into neural networks, because this coupling can improve the predictions significantly.
One of the first combinations of neural networks and constitutive modeling of solids are the works by \citet{shen2004} and \citet{liang2008}. Since then many different approaches were published, e.g.~physics-informed neural networks (PINNs) \citep{lagaris1998,raissi2019} with many extensions and variations, e.g.~\citep{kashefi2022,jagtap2020,henkes2022}. See also \citep{hu2024} for a recent review on developments of PINNs. \citet{linka2021} improved data-driven constitutive modeling by continuum-mechanics knowledge in a unified network approach referred to as constitutive artificical neural networks (CANNs). In \citep{linka2023} this concept was enhanced to autonomously discover constitutive models and learn an associated set of physically meaningful parameters. The variety of this research field furthermore covers physics-augmented neural networks \citep{liu2021, klein2023, fuhg2024b, benady2024}, mechanics-informed ones \citep{asad2022, li2023} and many more neural network approaches incorporating physics \citep{zlatic2024, pierre2023, canadija2024, chen2022, li2023, meyer2023, settgast2020}. Two contributions, \citep{linden2023} and \citep{klein2022}, will later serve as valuable references in Section \ref{sec_designing_iso_poly_functions} because of their role for polyconvex hyperelastic models. \citet{linden2023} proposed a comprehensive guide to enforce physics in neural networks for hyperelasticity and \citet{klein2022} provided two neural network frameworks for polyconvex anisotropic hyperelasticity. The present study will precisely add a new focus towards a more general incorporation of polyconvexity with a universal approximation property in isotropic hyperelasticity.

%%%%%%%%%%%%%%%%%%%%%%%%%%%%%%%%%%%%%%%%%%%
\subsection{Challenges of physical constraints for hyperelastic modeling} \label{sec_fundamentals}
The mathematical formulation of material models is subject to various principles and restrictions, of which this work focuses specifically on isotropic hyperelastic materials. From a physical viewpoint, a hyperelastic model should \citep{truesdell1965}:
\begin{enumerate}
\item Be path-independent. Accordingly the stresses derive from a potential $\Psi$, e.g.~the first Piola--Kirchhoff stress tensor
\begin{equation}
\boldsymbol P = \partial \Psi / \partial \boldsymbol F, \label{eq:definition_P}
\end{equation}
where $\boldsymbol F$ is the deformation gradient.
\item Fulfill the second law of thermodynamics (which trivially yields zero dissipation by the existence of a potential $\Psi$)
\begin{equation}
\boldsymbol P : \dot{\boldsymbol F} - \dot{\Psi} = 0.
\end{equation}
\item Be objective with respect to the orthogonal group of rotations $\SO(3)$ applied to the current configuration (frame-indifference)
\begin{align}
\Psi (\boldsymbol{F}) =\Psi (\boldsymbol Q \cdot \boldsymbol{F}) \quad \forall \, \boldsymbol Q \in \SO(3)
\label{eq:def:obejctiv}
\end{align}
\item and hence yields dependence of $\Psi$ on the right Cauchy--Green tensor $\boldsymbol C$ implicitly. As a consequence,
\begin{equation}
\boldsymbol P \cdot \boldsymbol{F}^T = 2 \, \boldsymbol{F} \cdot \frac{\partial \Psi}{\partial \boldsymbol{C}} \cdot \boldsymbol{F}^T = \boldsymbol{F} \cdot \boldsymbol P^T
\end{equation}
and thus balance of angular momentum is a priori fulfilled.
\item Be isotropic 
\begin{align}
\Psi (\boldsymbol{F}) =\Psi (\boldsymbol{F} \cdot \boldsymbol Q) \quad \forall \, \boldsymbol Q \in \SO(3)
\label{eq:def:isotropic}
\end{align}
\item and capture the growth conditions and avoid self penetration
\begin{subequations}
\begin{align}
&\Psi (\boldsymbol{F}) \rightarrow \infty \text{ as }\text{det}\,\boldsymbol{F} \rightarrow 0^+ \quad \text{and} \quad \Psi (\boldsymbol{F}) = \infty \text{ as } \det(\boldsymbol F) \leq 0\,, \label{eq:growthcond_a}\\
&\Psi (\boldsymbol{F}) \rightarrow \infty \text{ as }\lbrace \Vert \boldsymbol{F} \Vert + \Vert\text{cof}\,\boldsymbol{F}\Vert +\text{det}\,\boldsymbol{F}\rbrace \rightarrow \infty.\label{eq:growthcond_b}
\end{align}
\end{subequations}
\end{enumerate}
Principles like determinism, equipresence and local action in space and time are apparently also a priori fulfilled in this case. Further constraints such as energy and stress normalization can also be enforced if desired.

From a mathematical perspective one is interested in the lower semicontinuity of the energy
\begin{equation*}
		\boldsymbol u \mapsto I (\boldsymbol u) =\int_\Omega \Psi(\nabla \boldsymbol u)\, \mathrm{d}\omega
	\end{equation*}
since it ensures the existence of a minimizer in the set of suitable deformations $\boldsymbol u$. The foundational results of \citet{Mor52, Mor66} and subsequent refinements by \citet{Mey65}, \citet{AF84} and \citet{Mar85} show that the energy functional $I$ is lower semicontinuous and admits a minimizer if the energy $\Psi$ is quasiconvex and satisfies appropriate growth and coercivity conditions; see also \cite[Theorem 8.29]{Dac08}. However, the growth conditions imposed in these results preclude energy densities that attain the value $\infty$ and so conditions such as \eqref{eq:growthcond_a} and \eqref{eq:growthcond_b} cannot be satisfied. Consequently, the framework relying on the concept of quasiconvexity has limited applicability within the field of elasticity.

Following \citep{Bal76, ball1977}, the lower semicontinuity and existence of a minimizer for $I$ can be ensured under weaker growth and coercivity conditions by restricting to a smaller class of functions regarding convexity -- specifically, polyconvex functions; see also \cite[Theorem 8.31]{Dac08}. In particular, the mathematical framework for the notion of polyconvexity is compatible with the other principles and restrictions for hyperelastic materials. Although polyconvexity implies quasiconvexity, the framework based on polyconvexity is less restrictive with respect to physical constraints.

\begin{itemize}
	\item[7.] Therefore, $\Psi \colon \R^{3 \times 3} \to \R_\infty \coloneqq \R\, \cup\,\infty$ should be polyconvex, i.e.~there exists a convex and lower semicontinuous function $\widehat{\Psi} \colon \R^{3 \times 3} \times \R^{3 \times 3} \times \R \to  \R_\infty$ such that
	\begin{equation}\label{eq:def:Polyconvex}
		\Psi(\boldsymbol{F}) = \widehat{\Psi}(\boldsymbol{F},\, \text{cof}\,\boldsymbol{F},\, \text{det}\,\boldsymbol{F}) \text{ convex in } \boldsymbol{F},\, \text{cof}\,\boldsymbol{F} \,\text{ and } \text{det}\,\boldsymbol{F} \text{ for all } \boldsymbol{F}\in \R^{3 \times 3}\,.
	\end{equation}
\end{itemize} 

%%%%%%%%%%%%%%%%%%%%%%%%%%%%%%%%%%%%%%%%%%%
\subsection{Open challenge of a universal, frame-indifferent, isotropic polyconvex ANN approximation for hyperelasticity} 

The current aim is to develop a neural network framework for isotropic hyperelasticity that fulfills all desired physical and mathematical principles \eqref{eq:definition_P}--\eqref{eq:def:Polyconvex} in an exact way and still possesses a universal approximation capability. To achieve an integration of the physical and mathematical constraints without undesired restrictions of the solution space, the work develops along the following highlights:
\begin{itemize}
\item A sufficient and necessary criterion for frame-indifferent, isotropic polyconvex functions
\item A novel neural network framework strictly enforcing this criterion
\item Proof of a universal approximation theorem for frame-indifferent, isotropic polyconvex functions for the presented approach
\item Numerical examples in hyperelasticity demonstrating advantages for energies showing also linear growth in the stretch, non-convex energies and polyconvex hulls
\end{itemize}

Following this introductory section, Section~2 will point out the difficulties and currently established solutions regarding the modeling of frame-indifferent, isotropic polyconvex functions. This is done in the form of a brief comparison between two existing approaches and a third option based on a sufficient and necessary criterion for frame-indifferent, isotropic polyconvex functions \citep{wiedemann2023}. The latter will be chosen for the design of an improved framework in Section~3 that encompasses the class of Convex Signed Singular Value Neural Networks (CSSV-NNs). Section~4 provides a proof for a universal approximation theorem for these CSSV-NNs in the context of frame-indifferent, isotropic polyconvex functions. The algorithmic implementation, hyperparameters as well as details for the training and for the approximation of polyconvex hulls are described in Section~5. Finally, numerical results for different energies and the computation of a polyconvex hull are presented and discussed in Section~6 before the work is concluded in Section~7.

%%%%%%%%%%%%%%%%%%%%%%%%%%%%%%%%%%%%%%%%%%%
\section{Designing frame-indifferent and polyconvex functions} \label{sec_designing_iso_poly_functions}
%%%%%%%%%%%%%%%%%%%%%%%%%%%%%%%%%%%%%%%%%%%
The design of material models or their neural network surrogates becomes even more challenging when the aforementioned physical and mathematical constraints need to be considered. Especially the restrictions associated with frame-indifference, isotropy and polyconvexity are difficult to reconcile since naive approaches often lead to contradictions -- as will also be highlighted later. Several modeling frameworks were hence developed for this task. In the following, we recap two frameworks that are also used in the context of polyconvex material modeling with neural networks. Delivering a valuable basis, their different focus yet misses the exact fulfillment of frame-indifference or uses an overly strict implementation, respectively. 
While these two approaches will be briefly recapped as references, a third approach is discussed in more detail to derive an improved strategy for neural network modeling. Its advantage lies in a sufficient and necessary (and thus equivalent) condition for polyconvexity under the assumption of frame-indifference and isotropy.

%%%%%%%%%%%%%%%%%%%%%%%%%%%%%%%%%%%%%%%%%%%
\subsection{Approach 1: Anisotropic polyconvex hyperelasticity}
The first approach is for the more general, anisotropic case. \citet{klein2022} proposed to work directly with the definition of polyconvexity leading to parametrization
\begin{align}
\widehat{\Psi} (\boldsymbol{F}, \text{cof}\,\boldsymbol{F}, \text{det}\,\boldsymbol{F}), \label{eq_klein_energy}
\end{align}
where convexity of $\widehat{\Psi}$ is implemented by the neural network design. While the energy in~\eqref{eq_klein_energy} is by design polyconvex, it is not a priori frame-indifferent. Certainly, one could alternatively start directly from an energy in terms of $\boldsymbol C=\boldsymbol F^T \! \cdot \! \boldsymbol F$. In this case, frame-indifference would be a priori fulfilled. However, enforcing polyconvexity for a representation in terms of $\boldsymbol C$ remains unclear. For instance, the St.~Venant--Kirchhoff model is convex in $\boldsymbol C$ (and thus polyconvex in $\boldsymbol C$), but it is known to be not polyconvex in $\boldsymbol F$, cf.~\citep{raoult1986}. As a consequence, a rigorous a priori enforcement of frame-indifference and polyconvexity at the same time is not possible for the most general anisotropic case -- at least, not at the present time. For this reason, frame-indifference of energy~\eqref{eq_klein_energy} is only approximated in \citep{klein2022} on a discrete set of rotations. Similarly, invariance with respect to a given symmetry group can be enforced or approximated. This is implemented by data augmentation for the training of a neural network.

%%%%%%%%%%%%%%%%%%%%%%%%%%%%%%%%%%%%%%%%%%%
\subsection{Approach 2: Isotropic polyconvex hyperelasticity based on invariants}
\citet{linden2023} worked directly with invariants $\mathcal I_1 \!\coloneq \text{tr}\,\boldsymbol{C} = \boldsymbol F \! : \! \boldsymbol F$, $\mathcal I_2 \!\coloneq \text{tr}(\text{cof}\,\boldsymbol{C}) = \text{cof}\,\boldsymbol{F} \! : \! \text{cof}\,\boldsymbol{F}$ and $\mathcal I_3 \!\coloneq \sqrt{\text{det}\,\boldsymbol{C}} = \text{det}\,\boldsymbol{F}$ of the right Cauchy--Green tensor $\boldsymbol{C}$, a priori fulfilling frame-indifference and isotropy, i.e.~
\begin{align}
\Psi (\boldsymbol{F}) = \check\Psi(\mathcal I_1,\mathcal I_2,\mathcal I_3). \label{eq_linden_energy}
\end{align}
Since the invariants $\mathcal I_1$, $\mathcal I_2$ and $\mathcal I_3$ are evidently convex in $\boldsymbol F$, $\text{cof}\,\boldsymbol{F}$ and $\text{det}\,\boldsymbol{F}$, a sufficient criterion for $\Psi$ to be polyconvex is that $\check\Psi$ is convex and non-decreasing in the first two invariants and convex in the third. The non-decreasing constraint originates from the fact that the composition of a non-decreasing convex function with a convex function is convex. Since this is a sufficient, but not a necessary condition, some polyconvex energies cannot be designed by means of Eq.~\eqref{eq_linden_energy}. To be more precise, for instance, terms that are linear in $\boldsymbol{F}$ can not be approximated since that requires the application of the square root, which is not convexity preserving. Thus, working with invariants of the right Cauchy--Green tensor is not ideal.

%%%%%%%%%%%%%%%%%%%%%%%%%%%%%%%%%%%%%%%%%%%
\subsection{Approach 3: Isotropic polyconvex hyperelasticity based on signed singular values}
\citet{wiedemann2023} proposed to work with the signed singular values ($\nu_1$, $\nu_2$ and $\nu_3$) of $\boldsymbol F$ in order to obtain a sufficient and necessary condition for the polyconvexity of $\Psi$; see also \citep{dacorogna1993} for the 2D case. Starting from the spectral decomposition of the right Cauchy--Green deformation tensor
\begin{equation}
\boldsymbol C = \boldsymbol F^{T} \cdot \boldsymbol F = \sum_{i=1}^{3} \lambda_i^2 \, \boldsymbol{N}_i \otimes \boldsymbol{N}_i,
\end{equation}
the singular values $\sigma_i$ of $\boldsymbol{F}$ (principal stretches) are simply the square root of the eigenvalues of $\boldsymbol C$, i.e.~$\sigma_i = \sqrt{\lambda_i^2} > 0$. The signed singular values $\nu_i$ have the same absolute value, $|\nu_i| = \sigma_i$, but can become negative by rotation. The choice of their sign, however, does not affect the determinant (orientation preserving), yielding  $\nu_1 \, \nu_2 \, \nu_3 = \sigma_1 \, \sigma_2 \, \sigma_3$.
Apparently, frame-indifference can be a priori guaranteed by using (signed) singular values, as they relate to the frame-indifferent eigenvalues. 

For every frame-indifferent and isotropic (energy) function $\Psi(\boldsymbol F)$ there exists a unique function $\widetilde{\Psi}(\nu_1,\nu_2,\nu_3)$ that characterizes $\Psi$ in terms of the signed singular values of $\boldsymbol{F}$; see also \cite[Proposition 5.31]{Dac08}. This description will help to achieve polyconvexity. In line with \citep{wiedemann2023}, the function $\widetilde{\Psi}$ is referred to as singular value polyconvex if the associated $\Psi$ is polyconvex, i.e.~
\begin{equation}
\widetilde{\Psi} \text{ singular value polyconvex} :\Leftrightarrow \Psi \text{ polyconvex}
\end{equation}

As an intermediate step, we consider frame-indifference and isotropy, or mathematically speaking $SO(3) \times SO(3)$-invariance of $\Psi$. This implies that $\widetilde{\Psi}$ is $\Pi_3$-invariant with 
\begin{align}
\Pi_3 \coloneq \lbrace \boldsymbol B \cdot \textbf{diag}(\epsilon) \, \vert \, \boldsymbol B \in \text{Perm}(3), \, \epsilon \in \lbrace -1,1 \rbrace^3, \, \overset{\bullet}{\epsilon} = \epsilon_1 \, \epsilon_2 \, \epsilon_3 = 1 \rbrace,
\end{align} 
where Perm$(3)$ denotes the set of permutation matrices. To give a simplified interpretation, the $\Pi_3$-invariance includes the four symmetries
\begin{align}
\widetilde{\Psi} (\nu_1, \nu_2, \nu_3) = \widetilde{\Psi} (-\nu_1, -\nu_2, \nu_3) = \widetilde{\Psi} (-\nu_1, \nu_2, -\nu_3) = \widetilde{\Psi} (\nu_1, -\nu_2, -\nu_3)
\end{align}
together with the six invariances of $\widetilde{\Psi}$ with respect to permutations of the signed singular values, e.g. $\widetilde{\Psi} (\nu_1, \nu_2, \nu_3) = \widetilde{\Psi} (\nu_2, \nu_3, \nu_1)$ (see Tab.~\ref{tab:input_permutations} for all combinations). The condition $\epsilon_1 \, \epsilon_2 \, \epsilon_3 = 1$ enforces the local deformation (deformation gradient) to be orientation preserving. For the isotropic case, we can eventually approach polyconvexity by combining the signed singular values and the $\Pi_3$-invariance into the following equivalence. 

\begin{theorem}[\textbf{Singular value polyconvexity, cf.~\citep{wiedemann2023}}]\label{theo_singvalpolyconv}
A $\Pi_3$-invariant function $\widetilde{\Psi}\colon \, \R^3 \rightarrow \R_\infty$ is singular value polyconvex (and thus $\Psi$ is polyconvex) if and only if there exists a lower semicontinuous and convex function $\overline{\Psi}\colon \R^7 \rightarrow \R_\infty$ fulfilling
\begin{align}
\widetilde{\Psi} (\nu_1, \nu_2, \nu_3) = \overline{\Psi} (\nu_1, \nu_2, \nu_3, \nu_1 \nu_2, \nu_1 \nu_3, \nu_2 \nu_3, \nu_1 \nu_2 \nu_3) \quad \forall \, \boldsymbol \nu \in \R^3 \text{ with }\boldsymbol{\nu} = (\nu_1, \nu_2, \nu_3) \label{eq_psiisoandconv}
\end{align}
\end{theorem}
A proof of Theorem \ref{theo_singvalpolyconv} was given by \citet{wiedemann2023}. By defining the notation for the elementary polynomials
\begin{equation}
m(\boldsymbol{\nu}) = ( \nu_1, \nu_2, \nu_3, \nu_1 \nu_2, \nu_1 \nu_3, \nu_2 \nu_3, \nu_1 \nu_2 \nu_3 )
\label{eq:elementary_polynomials}
\end{equation}
one can evidently rewrite \eqref{eq_psiisoandconv} in the more compact form
\begin{align}
\widetilde{\Psi} = \overline{\Psi} \circ m.
\end{align}
It should be noted for the following that some
symbols may appear as functions or tensors, respectively. For brevity and readability, we omit a distinction by notation when
the context is clear. Otherwise, their role is explicitly noted.

In short, the polyconvexity criterion can be equivalently replaced in the case of frame-indifference and isotropy as
\begin{equation}
\begin{split}
&\Psi (\boldsymbol{F}) \text{ polyconvex} \\
\Leftrightarrow \,
&\widehat{\Psi} (\boldsymbol{F}, \text{cof}\,\boldsymbol{F}, \text{det}\,\boldsymbol{F}) \text{ convex}\ \\
\Leftrightarrow \,
&\widetilde{\Psi}(\nu_1, \nu_2, \nu_3) \text{ singular value polyconvex and $\Pi_3$-invariant} \label{eq:equivalence}\\
\Leftrightarrow \,
&\overline{\Psi}(\nu_1, \nu_2, \nu_3, \nu_1 \nu_2, \nu_1 \nu_3, \nu_2 \nu_3, \nu_1 \nu_2 \nu_3) \text{ convex and $\Pi_3$-invariant (w.r.t.~$\nu_1, \, \nu_2, \, \nu_3$)}.
\end{split}
\end{equation}
This combined sufficient and necessary condition for a frame-indifferent, isotropic polyconvex energy potential is a powerful tool for the formulation and determination of such energies and -- as will be shown in the next section -- suits neural network approaches perfectly. Note that it can also be employed to approximate the polyconvex hull of isotropic functions \citep{neumeier2024} with neural networks as will be shown in Section~\ref{sec_convexhull} or for numerical approximation schemes.

\begin{table}[h!]
\caption{Input permutations of the signed singular values for the neural network employed in Eq.~\eqref{nn_energy}.}
\renewcommand{\arraystretch}{2}
\resizebox{\textwidth}{!}{%
\begin{tabular}{|l|l|l|}
\hline
$\boldsymbol x^{(1)} = \left( \nu_1, \nu_2, \nu_3, \nu_1 \nu_2, \nu_1 \nu_3, \nu_2 \nu_3, \nu_1 \nu_2 \nu_3 \right)$ & $\boldsymbol x^{(2)} = \left( -\nu_1, -\nu_2, \nu_3, \nu_1 \nu_2, -\nu_1 \nu_3, -\nu_2 \nu_3, \nu_1 \nu_2 \nu_3 \right)$ & $\boldsymbol x^{(3)} = \left( -\nu_1, \nu_2, -\nu_3, -\nu_1 \nu_2, \nu_1 \nu_3, -\nu_2 \nu_3, \nu_1 \nu_2 \nu_3 \right)$ \\\hline
$\boldsymbol x^{(4)} = \left( \nu_1, -\nu_2, -\nu_3, -\nu_1 \nu_2, -\nu_1 \nu_3, \nu_2 \nu_3, \nu_1 \nu_2 \nu_3 \right)$  & $\boldsymbol x^{(5)} = \left( \nu_1, \nu_3, \nu_2, \nu_1 \nu_3, \nu_1 \nu_2, \nu_2 \nu_3, \nu_1 \nu_2 \nu_3 \right)$  & $\boldsymbol x^{(6)} = \left( -\nu_1, -\nu_3, \nu_2, \nu_1 \nu_3, -\nu_1 \nu_2, -\nu_2 \nu_3, \nu_1 \nu_2 \nu_3 \right)$ \\ \hline
$\boldsymbol x^{(7)} = \left( -\nu_1, \nu_3, -\nu_2, -\nu_1 \nu_3, \nu_1 \nu_2, -\nu_2 \nu_3, \nu_1 \nu_2 \nu_3 \right)$  & $\boldsymbol x^{(8)} = \left( \nu_1, -\nu_3, -\nu_2, -\nu_1 \nu_3, -\nu_1 \nu_2, \nu_2 \nu_3, \nu_1 \nu_2 \nu_3 \right)$   & $\boldsymbol x^{(9)} = \left( \nu_2, \nu_1, \nu_3, \nu_1 \nu_2, \nu_2 \nu_3, \nu_1 \nu_3, \nu_1 \nu_2 \nu_3 \right)$  \\ \hline
$\boldsymbol x^{(10)} = \left( -\nu_2, -\nu_1, \nu_3, \nu_1 \nu_2, \nu_2 \nu_3, \nu_1 \nu_3, \nu_1 \nu_2 \nu_3 \right)$ & $\boldsymbol x^{(11)} = \left( -\nu_2, \nu_1, -\nu_3, -\nu_1 \nu_2, \nu_2 \nu_3, -\nu_1 \nu_3, \nu_1 \nu_2 \nu_3 \right)$ & $\boldsymbol x^{(12)} = \left( \nu_2, -\nu_1, -\nu_3, -\nu_1 \nu_2, -\nu_2 \nu_3, \nu_1 \nu_3, \nu_1 \nu_2 \nu_3 \right)$ \\ \hline
$\boldsymbol x^{(13)} = \left( \nu_3, \nu_1, \nu_2, \nu_3 \nu_1, \nu_3 \nu_2, \nu_1 \nu_2, \nu_1 \nu_2 \nu_3 \right)$ & $\boldsymbol x^{(14)} = \left( -\nu_3, -\nu_1, \nu_2, \nu_3 \nu_1, -\nu_3 \nu_2, -\nu_1 \nu_2, \nu_1 \nu_2 \nu_3 \right)$ & $\boldsymbol x^{(15)} = \left( -\nu_3, \nu_1, -\nu_2, -\nu_3 \nu_1, \nu_3 \nu_2, -\nu_1 \nu_2, \nu_1 \nu_2 \nu_3 \right)$ \\ \hline
$\boldsymbol x^{(16)} = \left( \nu_3, -\nu_1, -\nu_2, -\nu_3 \nu_1, -\nu_3 \nu_2, \nu_1 \nu_2, \nu_1 \nu_2 \nu_3 \right)$ & $\boldsymbol x^{(17)} = \left( \nu_2, \nu_3, \nu_1, \nu_2 \nu_3, \nu_2 \nu_1, \nu_3 \nu_1, \nu_1 \nu_2 \nu_3 \right)$ & $\boldsymbol x^{(18)} = \left( -\nu_2, -\nu_3, \nu_1, \nu_2 \nu_3, -\nu_2 \nu_1, -\nu_3 \nu_1, \nu_1 \nu_2 \nu_3 \right)$ \\ \hline
$\boldsymbol x^{(19)} = \left( -\nu_2, \nu_3, -\nu_1, -\nu_2 \nu_3, \nu_2 \nu_1, -\nu_3 \nu_1, \nu_1 \nu_2 \nu_3 \right)$ & $\boldsymbol x^{(20)} = \left( \nu_2, -\nu_3, -\nu_1, -\nu_2 \nu_3, -\nu_2 \nu_1, \nu_3 \nu_1, \nu_1 \nu_2 \nu_3 \right)$ & $\boldsymbol x^{(21)} = \left( \nu_3, \nu_2, \nu_1, \nu_3 \nu_2, \nu_3 \nu_1, \nu_2 \nu_1, \nu_1 \nu_2 \nu_3 \right)$ \\ \hline
$\boldsymbol x^{(22)} = \left( -\nu_3, -\nu_2, \nu_1, \nu_3 \nu_2, -\nu_3 \nu_1, -\nu_2 \nu_1, \nu_1 \nu_2 \nu_3 \right)$ & $\boldsymbol x^{(23)} = \left( -\nu_3, \nu_2, -\nu_1, -\nu_3 \nu_2, \nu_3 \nu_1, -\nu_2 \nu_1, \nu_1 \nu_2 \nu_3 \right)$ & $\boldsymbol x^{(24)} = \left( \nu_3, -\nu_2, -\nu_1, -\nu_3 \nu_2, -\nu_3 \nu_1, \nu_2 \nu_1, \nu_1 \nu_2 \nu_3 \right)$ \\ \hline
\end{tabular}%
}
\label{tab:input_permutations}
\end{table}

\section{Physics-based convex signed singular value neural network for isotropic hyperelasticity} \label{chp_ann}

\begin{figure}[htp!]
\centering
\includegraphics[width=\textwidth]{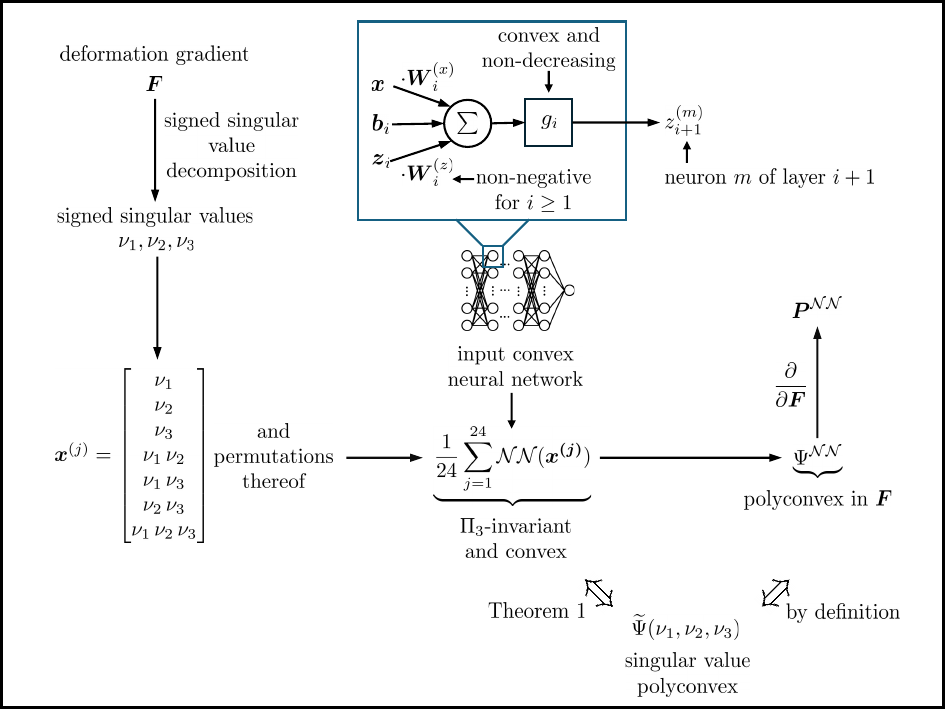}
\caption{Illustration of the convex signed singular value neural network (CSSV-NN) framework.}\label{fig_cssv_nn}
\end{figure}

Using Theorem \ref{theo_singvalpolyconv} (Eq.~\eqref{eq_psiisoandconv}), we present a new neural network framework that is able to approximate any frame-indifferent, isotropic polyconvex function. We achieve this by an input convex neural network (ICNN) \citep{amos2017} that works with the elementary polynomials of the signed singular values as input. In order to account for the $\Pi_3$-invariance of the potential, we propose
\begin{align}
\Psi^\nn &= \dfrac{1}{24} \sum_{j = 1}^{24} \nn(\boldsymbol x^{(j)}), \label{nn_energy}\\
\text{where} \quad \boldsymbol x^{(j)} & \in \text{six permutations times four reflections of } \boldsymbol x^{(1)}, \text{ cf.~Tab.~\ref{tab:input_permutations}}
\end{align}
See Tab.~\ref{tab:input_permutations} for all combinations of $\boldsymbol x^{(j)}$. It should be emphasized that the same $\nn$ and therefore the same weights have to be used for every input permutation in order to guarantee $\Pi_3$-invariance.
We define the neural network over the generalized input $\boldsymbol x$ using the architecture
\begin{align}
\boldsymbol z_{i+1} = g_i \left(\boldsymbol W_i^{(\boldsymbol z)} \cdot \boldsymbol z_i + \boldsymbol W_i^{(\boldsymbol x)} \cdot \boldsymbol x + \boldsymbol b_i \right) \text{ for } i = 0,...,k-1 \text{ and } \nn(\boldsymbol x;\theta) = \boldsymbol z_k, \label{eq:nn_architecture}
\end{align}
where $\boldsymbol z_i$ are the layer activations ($\boldsymbol z_0 \equiv 0,$ $\boldsymbol W_0^{(\boldsymbol z)} \equiv 0$), $\theta = \lbrace \boldsymbol W_{0:k-1}^{(\boldsymbol x)}, \boldsymbol W_{1:k-1}^{(\boldsymbol z)}, \boldsymbol b_{0:k-1} \rbrace$ are the weights, $g_i$ are non-linear activation functions and $k$ is the number of layers. The overall architecture is sketched in Fig.~\ref{fig_cssv_nn}. Then $\nn$ is convex in $\boldsymbol x$ if all $\boldsymbol W_{1:k-1}^{(\boldsymbol z)}$ are non-negative and all activation functions $g_i$ are convex and non-decreasing \citep{amos2017}. \citet{huang2021} and \citet{chen2019} provided a universal approximation theorem for convex functions for an architecture of type \eqref{eq:nn_architecture} when ReLu or Softplus activation functions are used, proving the representation power of ICNNs, see also Lemma \ref{lem:uat_conv_func}. A universal approximation theorem for the underlying neural network is proven in Section~\ref{sec_uat}.\\
To summarize, $\Psi^\nn$ represents a hyperelastic potential, is frame-indifferent, isotropic, polyconvex, guarantees a symmetric Cauchy stress $\boldsymbol \sigma$ and fulfills the Clausius--Duhem inequality. Hence, the framework fulfills all desired constraints 1--5 and 7 exactly. Further constraints, such as the growth conditions \eqref{eq:growthcond_a} and \eqref{eq:growthcond_b} or energy and stress normalization, can be easily enforced by adding respective terms to the potential
\begin{align}
\Psi = \Psi^\nn + \Psi^{\text{energy}} + \Psi^{\text{stress}} + \Psi^{\text{growth}}.
\end{align}
Since those constraints are not important for the underlying work and sometimes even not desired, they are not considered further at this point and the interested reader is referred to \citep{linden2023} for details on how to design such terms. Within this work those properties will be learned by the neural network from respective training data.\\
Finally, the first Piola--Kirchhoff stress tensor defined by the network potential reads
\begin{align}
\boldsymbol{P}^\nn = \dfrac{\partial \Psi^\nn}{\partial \boldsymbol F} = \dfrac{1}{24} \sum_{j = 1}^{24} \sum_{k = 1}^{3} \dfrac{\partial \nn(\boldsymbol x^{(j)})}{\partial \boldsymbol x^{(j)}} \cdot \dfrac{\partial \boldsymbol x^{(j)}}{\partial \nu_k} \, \dfrac{\partial \nu_k}{\partial \boldsymbol F}.
\end{align}

\begin{remark}
Working with invariants of $\boldsymbol C$ in a similar framework (compare \citep{linden2023}) requires all $\boldsymbol W_{0:k-1}^{(\boldsymbol x)}$ to be non-negative except for those associated with $\mathcal I_3$. This is quite restrictive and motivates the present improvement using signed singular values. Furthermore and as already mentioned, this corresponds to a sufficient but not necessary condition of polyconvexity.
\end{remark}

\clearpage
\section{Universal approximation theorem for frame-indifferent, isotropic polyconvex functions}\label{sec_uat}
In this section, we provide a universal approximation theorem for the frame-indifferent, isotropic polyconvex CSSV-NN, i.e.~we show that one can approximate any frame-indifferent, isotropic polyconvex function $\Psi$ up to arbitrary accuracy by a frame-indifferent, isotropic polyconvex CSSV-NN. First, we recall the universal approximation of convex functions by ICNNs. Second, we transfer this property to singular value polyconvex functions. The proof for frame-indifferent, isotropic polyconvex functions is completed by using the equivalence in \eqref{eq:equivalence}.

\begin{lemma}[\textbf{Universal approximation theorem for convex functions}]\label{lem:uat_conv_func}
Let 
%$a_i<b_i \in \R$ for $i \in \{1, \dots, n\}$  and 
$\Omega %= [a_1, b_1] \times \dots \times[a_n, b_n]
 \subset \R^n$ be a compact set and $f \colon \Omega \to \R$ be convex and, in particular,  continuous. Then, for every $\varepsilon >0$ there exists an ICNN such that
\begin{align}
\sup\limits_{\boldsymbol{x} \in \Omega}\, 
\left| \nn(\boldsymbol{x}) - f (\boldsymbol{x})\right| < \varepsilon\,.
\end{align}
\end{lemma}

\begin{proof}
The proposed network is an ICNN by design and a proof for the universal approximation theorem is provided in \citep{huang2021} for ReLU and Softplus activation functions for $\Omega = [0, 1]^n$. Their result can be easily extended to an arbitrary compact set $\Omega \subset \R^n.$%=[a_1, b_1] \times \dots \times[a_n, b_n]$ with $a_i,b_i \in \R$ and $b_i > a_i$. 
\end{proof}

\begin{theorem}[\textbf{Universal approximation theorem for singular value polyconvex functions}]\label{prop:uat-psi}
    
Let $\overline \Psi \colon \R^7 \to \R_\infty$ be lower semicontinuous and convex such that $\overline \Psi\circ \m$ is $\Pi_3$-invariant. Let $V\subset \R^3$ be a compact set where $\overline{\Psi}$ attains only finite values. Then, for every $\varepsilon > 0$, there exists a neural network $\Psi^\nn \colon \R^7 \rightarrow \R_\infty$ of the form \eqref{nn_energy} such that
\begin{align}\label{eq:poly-Approximation-result}
    \sup\limits_{\boldsymbol{\nu} \in V} |\Psi^\nn(m(\boldsymbol{\nu})) - \overline{\Psi}(m(\boldsymbol{\nu}))|
    &< \varepsilon\,.% && \forall \, \boldsymbol{\nu} \in V.
\end{align}
%where $V$ is the space characterized by finite energies. The more general case is considered in the appendix.
\end{theorem}

\begin{proof}
%This proof sketches the fundamental steps and a more detailed version is provided in the appendix.
Using Lemma \ref{lem:uat_conv_func}, we approximate $\overline{\Psi}$ on $\m(V)$ with $\nn$ by accuracy $\varepsilon$.
The $\Pi_3$-invariance of $\overline{\Psi} \circ \m$ implies $\overline{\Psi}(\m(\boldsymbol{\nu})) = \overline{\Psi}(\mathcal{P}_j (\m(\boldsymbol{\nu})))$ for $\boldsymbol x^{(j)} = \mathcal{P}_j (\boldsymbol x^{(1)})$ according to Tab.~\ref{tab:input_permutations}. Due to the group structure of $\Pi_3$ the equality holds also for $\mathcal{P}_j^{-1}$.
We compute with the $\triangle$-inequality, $\Pi_3$-invariance and the approximation by Lemma 2:

%The steps involved are: $\triangle$-inequality, $\Pi_3$-invariance and the group structure $\mathcal{G}$ of $\Pi_3$. The $\Pi_3$-invariance is defined by means of the permutation $\mathcal{P}_j$ that yields the input permutations of $x^{(1)}=m(\boldsymbol \nu)$ for the neural network as $\boldsymbol x^{(j)} = \mathcal{P}_j (\boldsymbol x^{(1)})$ according to Tab.~\ref{tab:input_permutations}.

%The steps involved are: $\triangle$-inequality, $\Pi_3$-invariance and the group structure $\mathcal{G}$ of $\Pi_3$. Furthermore, the $\Pi_3$-invariance is defined by means of the permutation $\mathcal{P}_j$ that yields the input permutations of $x^{(1)}=m(\boldsymbol \nu)$ for the neural network as $\boldsymbol x^{(j)} = \mathcal{P}_j (\boldsymbol x^{(1)})$ according to Tab.~\ref{tab:input_permutations}.

\begin{align*}
\sup\limits_{\boldsymbol{\nu} \in V} |\Psi^\nn(m(\boldsymbol{\nu})) - \overline{\Psi}(m(\boldsymbol{\nu}))|
=
&\sup\limits_{\boldsymbol{\nu} \in V} \dfrac{1}{24} \left| \sum_{j = 1}^{24} \left(\nn (\underbrace{\boldsymbol x^{(j)}}_{\mathcal{P}_j (\boldsymbol x^{(1)})}(\boldsymbol\nu)) - \overline{\Psi}(\m(\boldsymbol\nu))\right) \right| \\
\stackrel{\triangle}{\le}
&\dfrac{1}{24} \sum_{j = 1}^{24} \sup\limits_{\boldsymbol{\nu} \in V} \left| \left(\nn (\boldsymbol x^{(j)}(\boldsymbol\nu)) - \overline{\Psi}(\m(\boldsymbol\nu))\right) \right| \\
\stackrel{}{=}
&\dfrac{1}{24} \sum_{j = 1}^{24} \sup\limits_{\m(\boldsymbol{\nu}) \in \mathcal{P}_j(\m(V))}  \left| \left(\nn (m(\boldsymbol\nu)) - \overline{\Psi}(\mathcal{P}_j^{-1}(\m(\boldsymbol\nu)))\right) \right| \\
\stackrel{\Pi_3}{=}
&\dfrac{1}{24} \sum_{j = 1}^{24} \sup\limits_{\boldsymbol{\nu} \in V}  \left| \left(\nn (m(\boldsymbol\nu)) - \overline{\Psi}(\m(\boldsymbol\nu))\right) \right| \\
\stackrel{\text{Lemma 2}}{<}
& \dfrac{1}{24}  \sum_{j = 1}^{24} \varepsilon = \varepsilon\,.
\end{align*}
\end{proof}
The universal approximation for frame-indifferent, isotropic polyconvex functions now follows from combining Theorem \ref{prop:uat-psi} with the equivalence established in \eqref{eq:equivalence}.

\section{Algorithmic implementation} \label{sec_implementation}
The proposed neural network model was implemented using \textit{Python}	and \textit{TensorFlow}. 

\subsection{Network architecture} \label{sec_architecture}
The Softplus function $g^{SP}(x) = \text{ln}(1+\text{exp}(x))$ is employed as activation, except for the last layer, which has linear activation. Both functions are convex, non-decreasing and $C^\infty$ and thus fulfill the requirements for ICNNs used in the present polyconvex framework. Furthermore, this choice also guarantees universal approximation (see previous section). We directly set $\boldsymbol W_{k-1}^{(\boldsymbol x)} \equiv 0$, because the input arguments would cancel each other out due the employed symmetry anyway (except for the determinant for which the weight is also set to zero for implementational convenience). This does not affect the representation power or the universal approximation in any means.

Eight different network sizes were investigated in terms of number of layers and neurons per layer. We denote them by "size of input vector - number of neurons of hidden layer 1 - ... number of neurons of hidden layer $k-1$ - size of final layer $k$": 
\begin{multicols}{2}
\begin{enumerate}
\item 7-4-1 (36 parameters)
\item 7-8-1 (72 parameters)
\item 7-12-1 (108 parameters)
\item 7-4-2-1 (72 parameters)
\item 7-8-4-1 (160 parameters)
\item 7-12-8-1 (320 parameters)
\item 7-8-4-4-1 (236 parameters)
\item 7-12-8-4-1 (408 parameters)
\end{enumerate} 
\end{multicols} 
Recall that every hidden layer is also connected to the input via $\boldsymbol W_{i}^{(\boldsymbol x)}$. The above architectures were also motivated by the efficiency of deeper ICNNs as shown in \citet{chen2019} and chosen from preceding hyperparameter studies.

\subsection{Sobolev training}\label{sec_sobolev_training}
The different networks are trained in the Sobolev space by strain-stress tuples
\begin{align}
\mathcal{D} = \left\{ \left(\boldsymbol{F}^{(1)},\boldsymbol{P}^{(1)}\right), \, \left(\boldsymbol F^{(2)}, \boldsymbol{P}^{(2)}\right), \, ..., \, \left(\boldsymbol F^{(n)}, \boldsymbol{P}^{(n)}\right) \right\}
\end{align}
based on the mean squared error 
\begin{align}
\text{MSE} = \dfrac{1}{n} \sum_{i = 1}^{n} \left\Vert \boldsymbol{P}^{(i)}-\boldsymbol{P}^\nn \left(\boldsymbol{F}^{(i)}; \, \theta\right) \right\Vert^2 \label{eq_msestress}
\end{align}
with the Froebenius norm $\Vert \bullet \Vert$. The MSE represents deviations in the stress space and therefore even has a physical interpretation. Other choices for designing the loss function are of course also possible, e.g.~the networks can also be trained on strain-energy tuples or a combination of strain-energy and strain-stress tuples. However, training on strain-stress tuples only, yielded the best results for preliminary test cases and often is a natural choice when using experimental training data.\\
The following load cases have been used for training with analytical reference models (load direction $F_{ij}$, format $[$start value :  increment : end value$]$):
\begin{itemize}
\item no load $F_{ij} = \delta_{ij}$ (= $1$ if $i=j$, 0 else) %$\boldsymbol F = \boldsymbol I$
\item uniaxial compression/tension $F_{11} \in [0.9:0.02:1.1]\backslash \{ 1 \}$
\item large uniaxial compression $F_{11} \in [0.7:0.05:0.85]$
\item large uniaxial tension $F_{11} \in [1.5:0.5:10]$
\item biaxial compression/tension $F_{11} = F_{22} \in [0.9:0.02:1.1]\backslash \{ 1 \}$
\item volumetric compression/tension $F_{11} = F_{22} = F_{33} \in [0.9:0.02:1.1]\backslash \{ 1 \}$
\item simple shear $F_{12} \in [0.02:0.02:0.1]$
\end{itemize}
leading to 58 data points. The advantage of analytical reference models is that any number of data and load paths can be generated. Here, the training data set was selected in such a way that the training effort is appropriate, important aspects of the reference models are taken into account and the neural network learns the material behavior as holistic as possible. A validation data set is omitted for the noise-free analytical data base that is directly extracted from the ground truth reference. This does not limit any of the following statements. 

Finally, the weights $\theta$ are optimized with \textit{TensorFlow's} SGD optimizer in two steps, first with a learning rate of 0.001 and batch size of 2 for 1000 epochs, followed by a learning rate of 0.0001 and batch size of 2 for another 1000 epochs. 30 randomized weight initializations were investigated for each architecture and energy combination. 

\subsection{Approximation of polyconvex hulls} \label{sec_convexhull}
The presented framework can also be used for the approximation of a polyconvex hull defined as
\begin{align}
\Psi^{\text{pc}}(\boldsymbol{F}) \coloneq \text{sup}\, \{\mathcal{V}(\boldsymbol{F}) \, \vert \, \mathcal{V} \text{ polyconvex, } \mathcal{V} \leq \Psi \}
\end{align}
for a (potentially non-polyconvex) energy $\Psi$. Again, there exist multiple options on how to design a loss function for that case. The idea here is to enhance the loss function by a penalty term, which penalizes energies larger than the true energy to ensure
\begin{align}
\Psi^\nn \left(\boldsymbol{F}^{(i)}; \, \theta\right) \leq \Psi^{(i)} \quad \forall \, i \label{eq_smaller_energy}
\end{align}
for all data points in the set of strain-energy tuples
\begin{align}
\mathcal{D} = \left\{ \left(\boldsymbol{F}^{(1)},\Psi^{(1)}\right), \, \left(\boldsymbol F^{(2)}, \Psi^{(2)}\right), \, ..., \, \left(\boldsymbol F^{(n)}, \Psi^{(n)}\right) \right\}.
\end{align}
This is realized by the penalty function
\begin{align}
\text{PEN} = \alpha \, \sum_{i=1}^n \text{max}\left(\Psi^\nn \left(\boldsymbol{F}^{(i)}; \, \theta\right) - \Psi^{(i)}, 0 \right), 
\end{align} 
where $\alpha$ is a penalty factor. For the numerical example in Section~\ref{sec_num_approx_hull} a value of $\alpha = 1$ was already sufficient to ensure constraint \eqref{eq_smaller_energy}. The network is then trained based on the strain-energy tupels together with the penalty term, leading to the loss function
\begin{align}
\mathcal{L} = \dfrac{1}{n} \sum_{i = 1}^{n} \left\Vert \Psi^{(i)}-\Psi^\nn \left(\boldsymbol{F}^{(i)}; \, \theta\right) \right\Vert^2 + \text{PEN}. \label{eq_loss_hull}
\end{align}
Note that the polyconvex hull is only approximated in the domain of the training data. To be more precise, the predicted energy hull is polyconvex and smaller or equal to the provided energy within the considered data set.
\begin{remark}
Often, the convex hull shows regions where the energy grows linearly with respect to the stretch. This property once again underlines the importance of a universal approximation, which is achieved in this novel framework by using the sufficient and necessary criterion for polyconvexity based on signed singular values. By way of contrast an input convex neural network formulated in invariants of $\boldsymbol{C}$ such as in \citep{linden2023} can not capture such a linear growth (except for volumetric deformations).
\end{remark}

\section{Numerical results} \label{sec_results}
To validate the framework from Section~\ref{sec_implementation}, three different numerical examples are conducted. The first two examples focus on the identification of a hyperelastic energy for a given set of data following the implementation presented in Sections~\ref{sec_architecture} and \ref{sec_sobolev_training}, while the third one illustrates the ability to approximate polyconvex hulls according to Section~\ref{sec_convexhull}.\\
All neural network energy predictions were normalized for the first two examples such that $\Psi(\boldsymbol F = \boldsymbol I) = 0$ holds. As a helpful reference to explore the potentials and limitations of neural network modeling, we compare the results of the proposed CSSV-NN approach with predictions obtained from the neural networks of \citet{linden2023}, which are referred to as PANNs. The hyperparameter settings and the training procedure from Section~\ref{sec_sobolev_training} are equal for both approaches for comparison reasons. 

\subsection{CSSV-NN model of a polyconvex signed singular value energy}
\begin{figure}[htp!]
\centering
\includegraphics[width=\textwidth]{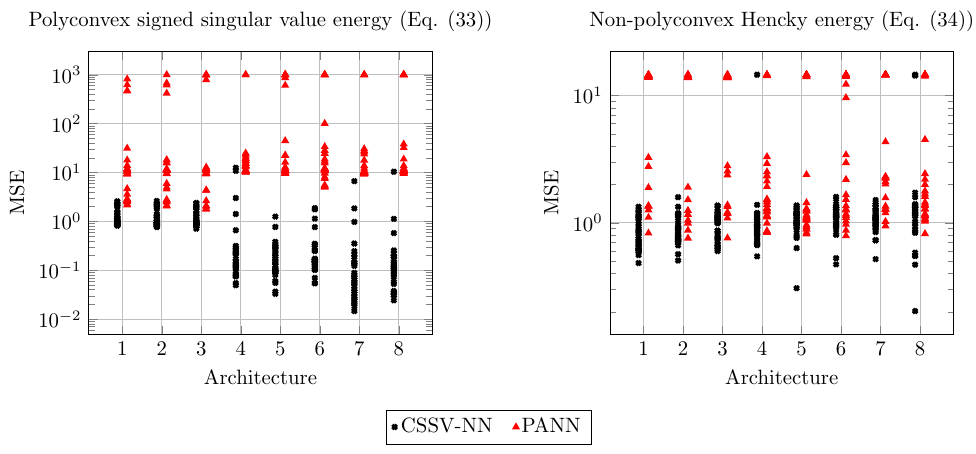}
\caption{Mean squared error of the training data set from each architecture and random initializations for the polyconvex signed singular value energy (left, Eq.~\eqref{eq:ssvenergy}) and the non-polyconvex Hencky energy (right, Eq.~\eqref{eq:henckyenergy}) after training.}\label{fig_loss_comparison}
\end{figure}

The first example is a polyconvex energy directly based on signed singular values
\begin{align}
\Psi^\ssve(\boldsymbol F) = \vert \nu_1(\boldsymbol F) \vert + \vert \nu_2(\boldsymbol F) \vert + \vert \nu_3(\boldsymbol F) \vert + \dfrac{1}{10 \, (\text{det}\,\boldsymbol{F})^{10}}. \label{eq:ssvenergy}
\end{align}
A hyperparameter study showed that architecture size 7 performs best in terms of the lowest overall MSE (as defined in Eq.~\eqref{eq_msestress}) for the training data set with a value of 0.0146, see Fig.~\ref{fig_loss_comparison} left. While the other large architectures (4, 5, 6 and 8) fall into approximately the same low regime of MSE values, the smaller models (1, 2 and 3) do not approximate the energy as accurately and show approximately 50 times higher errors for their best MSE. In contrast, none of the PANN models can capture the energy accurately. For every architecture the PANN framework performs worse than the CSSV-NNs as compared in Fig.~\ref{fig_loss_comparison}.
The best PANN model obtained for architecture 3 has a 140 times higher MSE than the best CSSV-NN. These differences can be explained by a closer look at the energy and stress responses.

\begin{figure}[htp!]
\centering
\caption*{\textbf{Polyconvex signed singular value energy (Eq.~\ref{eq:ssvenergy})}}
\includegraphics[width=\textwidth]{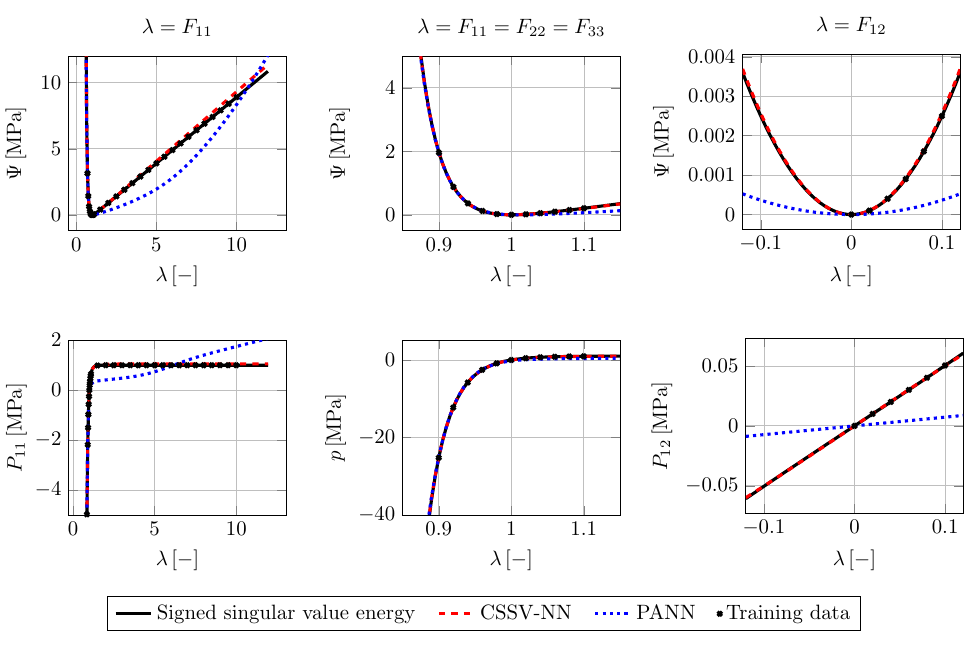}
\caption{Polyconvex signed singular value energy (Eq.~\eqref{eq:ssvenergy}): Energy and stress predictions from the CSSV-NN and the PANN in comparison with the ground-truth model for uniaxial stretch (left), volumetric expansion (center) and simple shear (right).}
\label{fig_sve_comparison}
\end{figure}

Considering the energy and stress predictions, we compare the best CSSV-NN architecture (size 7), the analytical reference and the best PANN architecture (size 3) for different load cases, see Fig.~\ref{fig_sve_comparison}.
The CSSV-NN fits the reference solution visually perfectly for volumetric compression/extension (middle of Fig.~\ref{fig_sve_comparison}) and simple shear (right). Even the extrapolation outside of the volumetric training stretches $0.9 \le \lambda \le 1.1$ and trained shear $0 \le \lambda \le 0.1$ works extremely well. Under uniaxial compression/tension (left), the prediction is perfect for smaller deformations, while small deviations occur for large tension. If better approximations are desired for specific deformations ranges, these can be accomplished by heavier weights in the loss function or more training data points in that region. We refrain from such a posteriori modifications, though, for better comparison and interpretation of the frameworks themselves. However, it should be emphasized once more that the minor differences can indeed be completely eliminated -- thanks to the universal approximation property of the neural network.

As already indicated by the MSE, the PANN model does not capture the model behavior fully. While the compression states are approximated well, the model fails for tension and simple shear. The reason is that the analytical model shows regimes where the energy grows linearly in the stretches of $\boldsymbol F$, an effect the PANN can not capture -- their lowest growth rate is quadratic except for volumetric compression/expansion. The parametrization chosen for the present CSSV-NN is also more beneficial for simple shear, since the gradients of the smaller shear stresses are negligible when compared to the high error sensitivity of the PANN in the other deformation states.

\subsection{CSSV-NN model of a non-polyconvex Hencky energy}
The second example is a non-polyconvex Hencky energy 
\begin{align}
\Psi^\hen(\boldsymbol F) = \dfrac{1}{2} \left( \text{ln}(\lambda_1) + \text{ln}(\lambda_2) + \text{ln}(\lambda_3) \right)^2 + \text{ln}(\lambda_1)^2 + \text{ln}(\lambda_2)^2 + \text{ln}(\lambda_3)^2 \label{eq:henckyenergy}
\end{align}
with $\lambda_i$ being the eigenvalues of the right stretch tensor $\boldsymbol U$ that are equal to the absolute values of the signed singular values of $\boldsymbol{F}$ (principal stretches). The material parameters in terms of the Lam\'e constants are set to $\lambda = 1 \, \text{MPa}$ and $\mu = 1 \, \text{MPa}$ corresponding to a Young's Modulus of $E = 2.5 \, \text{MPa}$ and a Poisson ratio of $\nu = 0.25$. According to \citep{bruhns2001}, this energy is not polyconvex. However, it is clear that both approaches under investigation, CSSV-NNs and PANNs, are inherently polyconvex and thus can not fit the non-polyconvex Hencky energy perfectly. However, it is often of great importance to find the best polyconvex approximation to non-polyconvex (training) data. 

The best CSSV-NN architecture (size 8) performs with a training data MSE of 0.203 (see Fig.~\ref{fig_loss_comparison} right).
\begin{figure}[htp!]
\centering
\caption*{\textbf{Non-polyconvex Hencky energy (Eq.~\ref{eq:henckyenergy})}}
\includegraphics[width=\textwidth]{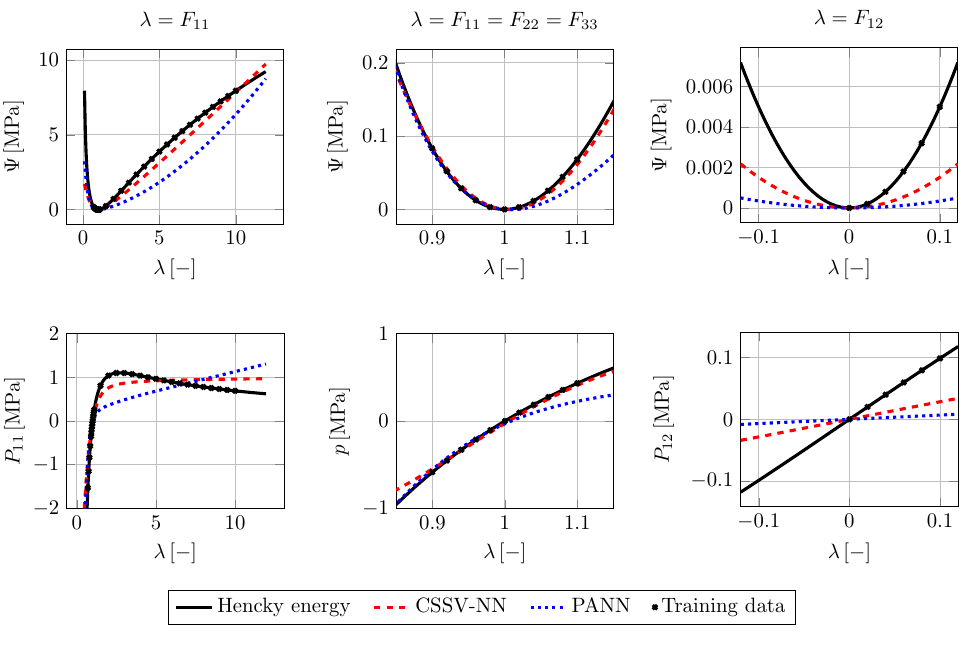}
\caption{Non-polyconvex Hencky energy 
(Eq.~\eqref{eq:henckyenergy}): Energy and stress predictions from the CSSV-NN and the PANN in comparison with the ground-truth model for uniaxial stretch (left), volumetric expansion (center) and simple shear (right).}
\label{fig_hencky_comparison}
\end{figure}
Again, it shows lower MSEs when compared to PANNs throughout all network sizes. This is due to the fact that the polyconvexification of an energy requires linear energy growth in terms of the stretches of $\boldsymbol F$ as already addressed above. The energy and stress predictions of the best architectures are depicted in Fig.~\ref{fig_hencky_comparison} and compared to the analytical reference solution. The CSSV-NN is able to reproduce the volumetric compression and tension behavior of the Hencky model well (middle). 

The most interesting part is uniaxial tension (Fig.~\ref{fig_hencky_comparison} left), where the Hencky model is non-(poly)convex. The CSSV-NN shows the best possible behavior for an inherently polyconvex surrogate by predicting constant stresses and a linear energy in the originally non-convex region. Again, the PANN is only able to predict quadratic energy growth. The training of both networks suffers from a high loss and small training gradients for the simple shear states, for which the CSSV-NN still provides a better quantitative fit (Fig.~\ref{fig_hencky_comparison} right).

\subsection{Approximation of a polyconvex hull}\label{sec_num_approx_hull}
In order to analyze the CSSV-NN's capability to approximate polyconvex hulls, we consider a model for the macroscopic response of nematic elastomers \citep{bladon1993}. It is examined here, because \citet{desimone2002} derived an analytic polyconvex hull for the proposed energy class that can be used as a reference solution. The (non-polyconvex) energy then reads
\begin{align}
\Psi^\nem(\boldsymbol F) = \begin{cases}
\dfrac{\lambda_1^p}{\gamma_1^p} + \dfrac{\lambda_2^p}{\gamma_2^p} + \dfrac{\lambda_3^p}{\gamma_3^p} - 3 & \text{if } \text{det}\,\boldsymbol{F}=1, \\
\infty & \text{else}, 
\end{cases} \label{eq:nematic energy}
\end{align}
with the already introduced eigenvalues $\lambda_1 \leq \lambda_2 \leq \lambda_3$ of the right stretch tensor $\boldsymbol U$ being equal to the absolute values of the signed singular values of $\boldsymbol{F}$ (principal stretches), the exponent $p \in \left(2,\infty\right)$ and the parameters $0<\gamma_1 \leq \gamma_2 \leq \gamma_3$ satisfying $\gamma_1 \, \gamma_2 \, \gamma_3 = 1$. Here, we choose $p=2$,  $\gamma_1 = 0.5$, $\gamma_2 = 1$ and $\gamma_3 = 2$. The infinite energy for deformations with $\text{det}\,\boldsymbol{F} \neq 1$ accounts for the incompressibility of nematic elastomers. We however do not have to consider this for the underlying application, since we only train and evaluate the neural network on data points of isochoric deformations. As a direct result, the energy only depends on two independent eigenvalues and the third follows from the constraint $\text{det}\,\boldsymbol{F} = 1$. The first two eigenvalues are discretized as $\lambda_1,\, \lambda_2 \in \Omega_d= \left[0.4,0.5,...,1,1.4,...,5\right]^2$ for training data generation. The energy of the training data is furthermore min--max normalized and shifted by $+0.5$ to improve training performance. Again, 30 randomized weight initializations were investigated and every network was trained with \textit{TensorFlow's} SGD optimizer for 1000 epochs with a learning rate of 0.1 and a batch size of 2 based on the penalty enhanced loss function given by Eq.~\eqref{eq_loss_hull}. The lowest overall loss for the training data domain is obtained for architecture 4 with a value of 0.00123. Note that it can never reach zero for a non-polyconvex energy due to the enforcement of polyconvexity.

\begin{figure}[htp!]
\centering
\caption*{\textbf{Approximation of a polyconvex hull (Eq.~\ref{eq:nematic energy})}}
\includegraphics[width=\textwidth]{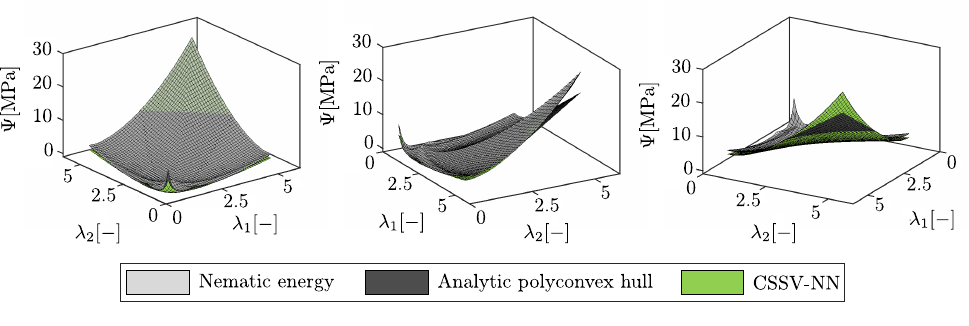}
\caption{Approximation of a polyconvex hull ((Eq.~\eqref{eq:nematic energy}): Comparison of the ground-truth (non-polyconvex) nematic energy, the analytic polyconvex hull and the CSSV-NN polyconvex hull in the principal stretch space.}\label{fig_nematic_comparison}
\end{figure}

The (non-polyconvex) nematic energy, the analytic polyconvex hull and the CSSV-NN convex hull prediction are depicted in Fig.~\ref{fig_nematic_comparison} for qualitative comparison.
While providing a polyconvex approximation that performs especially well at larger stretches, it tends to only slightly undershoot at three corners of the domain.
For a quantitative comparison, though, the domain of the polyconvexified hull must be considered. At first, one might expect that the network converges against the analytic polyconvex hull, but this is not the case. The reason is that the network only approximates the hull within the given domain of the training data, while the analytic polyconvexification considers the full domain $(0,\infty]^2$. Thus, the neural network solution is domain-dependent and the analytic polyconvex hull is domain-independent by construction. Depending on the intended application, this can be a major advantage if a polyconvex hull is only of interest for a limited range of operational loads. A domain-limited polyconvexification can then yield improved performance and solutions. At the same time, this can be a limitation of data-based frameworks operating on a bounded domain and requires a clear problem definition in practice. For the present training data domain, the error of the CSSV-NN network is smaller than the analytic approach, cf.~Fig.~\ref{fig_nematic_comparison}. For a finer discretization such as $\Omega_{\text{test}}= \left[0.4,0.41,0.42,...,5\right]^2$, for instance, the neural network achieves a loss value (according to Eq.~\eqref{eq_loss_hull} and without data normalization) of 1.23 while that of the analytic polyconvex hull is approximately $30 \, \%$ higher.

\clearpage
\section{Conclusion}

A novel Convex Signed Singular Value Neural Network (CSSV-NN) suitable for approximating isotropic hyperelastic energies was presented. It a priori fulfills frame-indifference, isotropy and polyconvexity while still satisfying the universal approximation theorem. Accordingly and in contrast to previous networks, any frame-indifferent, isotropic polyconvex hyperelastic energy can be approximated up to a desired accuracy. Its construction is based on the coupling of an input convex neural network with the elementary polynomials of the signed singular values as input. Respective invariances are ensured by a summation over network evaluations. A relevant, underlying property is the sufficient and necessary criterion for frame-indifferent, isotropic polyconvex functions. 
Practicability of the ANN implementation is guaranteed by building upon standard methods and constraints from \textit{Tensorflow}.

Compared to previous networks and due to the universal approximation theorem, the CSSV-NN showed superior performance for polyconvex and non-polyconvex energies. A particular advantage is the coverage of regimes where the energy shows linear growth with respect to the stretches. This property is especially important for approximating non-polyconvex energies and polyconvex hulls. Within the implementation for polyconvex hulls, the network only approximates the hull within the given domain of the training data, while the analytic polyconvexification considers the full domain. Thus, the neural network solution is domain-dependent and the analytic polyconvex hull is domain-independent by construction. Depending on the intended application, this can be a major advantage of the CSSV-NN. To be more precise, a domain-limited polyconvexification can yield an improved approximation within the considered data interval.

Employing signed singular values into an artificial neural network is indeed very powerful for approximating polyconvex energies. In future works, it shall be extended on the one hand to other material classes such as incompressible and anisotropic models. On the other hand, a hyperparameter guideline for experimental data is to be developed. 

\section*{Acknowledgements}
Financial support from the German Research Foundation (DFG) via SFB/TRR 188 (278868966), project C01, is gratefully acknowledged. Furthermore, the authors gratefully acknowledge the computing time
provided on the Linux HPC cluster at Technical University Dortmund (LiDO3), partially funded in the course of the Large-Scale Equipment Initiative by the German Research Foundation (DFG) as project 271512359.

\bibliographystyle{apalike-ejor}%
\bibliography{Literatur}%

\begin{thebibliography}{}

\bibitem[Acerbi \& Fusco, 1984]{AF84}
Acerbi, E. \& Fusco, N. (1984).
\newblock Semicontinuity problems in the calculus of variations.
\newblock {\em Arch. Rational Mech. Anal.}, 86(2), 125--145.
\newblock \url{https://doi.org/10.1007/BF00275731}

\bibitem[Amos et~al., 2017]{amos2017}
Amos, B., Xu, L., \& Kolter, J.~Z. (2017).
\newblock Input convex neural networks.
\newblock {\em Proceedings of the 34th International Conference on Machine
  Learning}, volume~70 of {\em Proceedings of Machine Learning Research},
  146--155.
\newblock \url{https://proceedings.mlr.press/v70/amos17b.html}

\bibitem[As'ad et~al., 2022]{asad2022}
As'ad, F., Avery, P., \& Farhat, C. (2022).
\newblock A mechanics-informed artificial neural network approach in
  data-driven constitutive modeling.
\newblock {\em International Journal for Numerical Methods in Engineering},
  123(12), 2738--2759.
\newblock \url{https://doi.org/https://doi.org/10.1002/nme.6957}

\bibitem[Ball, 1976]{Bal76}
Ball, J.~M. (1976).
\newblock On the calculus of variations and sequentially weakly continuous
  maps.
\newblock {\em Ordinary and partial differential equations ({P}roc. {F}ourth
  {C}onf., {U}niv. {D}undee, {D}undee, 1976)}, volume Vol. 564 of {\em Lecture
  Notes in Math.}, 13--25. Springer, Berlin-New York.

\bibitem[Ball, 1977]{ball1977}
Ball, J.~M. (1977).
\newblock Constitutive inequalities and existence theorems in nonlinear
  elastostatics.
\newblock {\em Nonlinear analysis and mechanics: Heriot-Watt symposium},
  volume~1, 187--241.

\bibitem[Bartel et~al., 2023]{bartel2023}
Bartel, T., Harnisch, M., Schweizer, B., \& Menzel, A. (2023).
\newblock A data-driven approach for plasticity using history surrogates:
  Theory and application in the context of truss structures.
\newblock {\em Computer Methods in Applied Mechanics and Engineering}, 414,
  116138.
\newblock \url{https://doi.org/https://doi.org/10.1016/j.cma.2023.116138}

\bibitem[Benady et~al., 2024]{benady2024}
Benady, A., Baranger, E., \& Chamoin, L. (2024).
\newblock Nn-mcre: A modified constitutive relation error framework for
  unsupervised learning of nonlinear state laws with physics-augmented neural
  networks.
\newblock {\em International Journal for Numerical Methods in Engineering},
  125(8), e7439.
\newblock \url{https://doi.org/https://doi.org/10.1002/nme.7439}

\bibitem[Bladon et~al., 1993]{bladon1993}
Bladon, P., Terentjev, E.~M., \& Warner, M. (1993).
\newblock Transitions and instabilities in liquid crystal elastomers.
\newblock {\em Phys. Rev. E}, 47, R3838--R3840.
\newblock \url{https://doi.org/10.1103/PhysRevE.47.R3838}

\bibitem[Bruhns et~al., 2001]{bruhns2001}
Bruhns, O.~T., Xiao, H., \& Meyers, A. (2001).
\newblock Constitutive inequalities for an isotropic elastic strain-energy
  function based on hencky's logarithmic strain tensor.
\newblock {\em Proceedings: Mathematical, Physical and Engineering Sciences},
  457(2013), 2207--2226.
\newblock \url{https://doi.org/http://doi.org/10.1098/rspa.2001.0818}

\bibitem[Canadija et~al., 2024]{canadija2024}
Canadija, M., Kosmerl, V., Zlatic, M., Vrtovsnik, D., \& Munjas, N. (2024).
\newblock A computational framework for nanotrusses: Input convex neural
  networks approach.
\newblock {\em European Journal of Mechanics - A/Solids}, 103, 105195.
\newblock
  \url{https://doi.org/https://doi.org/10.1016/j.euromechsol.2023.105195}

\bibitem[Chen \& Guilleminot, 2022]{chen2022}
Chen, P. \& Guilleminot, J. (2022).
\newblock Polyconvex neural networks for hyperelastic constitutive models: A
  rectification approach.
\newblock {\em Mechanics Research Communications}, 125, 103993.
\newblock
  \url{https://doi.org/https://doi.org/10.1016/j.mechrescom.2022.103993}

\bibitem[Chen et~al., 2019]{chen2019}
Chen, Y., Shi, Y., \& Zhang, B. (2019).
\newblock Optimal control via neural networks: A convex approach.
\newblock \url{https://arxiv.org/abs/1805.11835}

\bibitem[Dacorogna, 2008]{Dac08}
Dacorogna, B. (2008).
\newblock {\em Direct methods in the calculus of variations} (second ed.),
  volume~78 of {\em Applied Mathematical Sciences}.
\newblock Springer, New York.

\bibitem[Dacorogna \& Koshigoe, 1993]{dacorogna1993}
Dacorogna, B. \& Koshigoe, H. (1993).
\newblock On the different notions of convexity for rotationally invariant
  functions.
\newblock {\em Annales de la Facult\'e des sciences de Toulouse :
  Math\'ematiques}, Ser. 6, 2(2), 163--184.
\newblock \url{http://www.numdam.org/item/AFST_1993_6_2_2_163_0/}

\bibitem[DeSimone \& Dolzmann, 2002]{desimone2002}
DeSimone, A. \& Dolzmann, G. (2002).
\newblock Macroscopic {Response} of {Nematic} {Elastomers} via {Relaxation} of
  a {Class} of {SO}(3)-{Invariant} {Energies}.
\newblock {\em Archive for Rational Mechanics and Analysis}, 161(3), 181--204.
\newblock \url{https://doi.org/10.1007/s002050100174}

\bibitem[Fuhg et~al., 2024]{fuhg2024b}
Fuhg, J.~N., Jones, R.~E., \& Bouklas, N. (2024).
\newblock Extreme sparsification of physics-augmented neural networks for
  interpretable model discovery in mechanics.
\newblock {\em Computer Methods in Applied Mechanics and Engineering}, 426,
  116973.
\newblock \url{https://doi.org/https://doi.org/10.1016/j.cma.2024.116973}

\bibitem[Geuken et~al., 2024]{geuken2024}
Geuken, G.-L., Mosler, J., \& Kurzeja, P. (2024).
\newblock Incorporating sufficient physical information into artificial neural
  networks: A guaranteed improvement via physics-based rao-blackwellization.
\newblock {\em Computer Methods in Applied Mechanics and Engineering}, 423,
  116848.
\newblock \url{https://doi.org/https://doi.org/10.1016/j.cma.2024.116848}

\bibitem[Henkes et~al., 2022]{henkes2022}
Henkes, A., Wessels, H., \& Mahnken, R. (2022).
\newblock Physics informed neural networks for continuum micromechanics.
\newblock {\em Computer Methods in Applied Mechanics and Engineering}, 393,
  114790.
\newblock \url{https://doi.org/10.1016/j.cma.2022.114790}

\bibitem[Hu et~al., 2024]{hu2024}
Hu, H., Qi, L., \& Chao, X. (2024).
\newblock Physics-informed neural networks (pinn) for computational solid
  mechanics: Numerical frameworks and applications.
\newblock {\em Thin-Walled Structures}, 205, 112495.
\newblock \url{https://doi.org/https://doi.org/10.1016/j.tws.2024.112495}

\bibitem[Huang et~al., 2020]{huang2021}
Huang, C., Chen, R. T.~Q., Tsirigotis, C., \& Courville, A.~C. (2020).
\newblock Convex potential flows: Universal probability distributions with
  optimal transport and convex optimization.
\newblock {\em CoRR}, abs/2012.05942.
\newblock \url{https://arxiv.org/abs/2012.05942}

\bibitem[Jagtap \& Karniadakis, 2020]{jagtap2020}
Jagtap, A. \& Karniadakis, G. (2020).
\newblock Extended physics-informed neural networks (xpinns): A generalized
  space-time domain decomposition based deep learning framework for nonlinear
  partial differential equations.
\newblock {\em Communications in Computational Physics}, 28, 2002--2041.
\newblock \url{https://doi.org/10.4208/cicp.OA-2020-0164}

\bibitem[Kashefi \& Mukerji, 2022]{kashefi2022}
Kashefi, A. \& Mukerji, T. (2022).
\newblock Physics-informed pointnet: A deep learning solver for steady-state
  incompressible flows and thermal fields on multiple sets of irregular
  geometries.
\newblock {\em Journal of Computational Physics}, 468, 111510.
\newblock \url{https://doi.org/10.1016/j.jcp.2022.111510}

\bibitem[Kirchdoerfer \& Ortiz, 2016]{kirchdoerfer2016}
Kirchdoerfer, T. \& Ortiz, M. (2016).
\newblock Data-driven computational mechanics.
\newblock {\em Computer Methods in Applied Mechanics and Engineering}, 304,
  81--101.
\newblock \url{https://doi.org/10.1016/j.cma.2016.02.001}

\bibitem[Klein et~al., 2022]{klein2022}
Klein, D.~K., Fernández, M., Martin, R.~J., Neff, P., \& Weeger, O. (2022).
\newblock Polyconvex anisotropic hyperelasticity with neural networks.
\newblock {\em Journal of the Mechanics and Physics of Solids}, 159, 104703.
\newblock \url{https://doi.org/10.1016/j.jmps.2021.104703}

\bibitem[Klein et~al., 2023]{klein2023}
Klein, D.~K., Roth, F.~J., Valizadeh, I., \& Weeger, O. (2023).
\newblock Parametrized polyconvex hyperelasticity with physics-augmented neural
  networks.
\newblock {\em Data-Centric Engineering}, 4, e25.
\newblock \url{https://doi.org/10.1017/dce.2023.21}

\bibitem[Kumar \& Kochmann, 2022]{kumar2022}
Kumar, S. \& Kochmann, D.~M. (2022).
\newblock {\em What Machine Learning Can Do for Computational Solid Mechanics},
  275--285.
\newblock Springer International Publishing.
\newblock \url{https://doi.org/10.1007/978-3-030-87312-7_27}

\bibitem[Lagaris et~al., 1998]{lagaris1998}
Lagaris, I., Likas, A., \& Fotiadis, D. (1998).
\newblock Artificial neural networks for solving ordinary and partial
  differential equations.
\newblock {\em IEEE Transactions on Neural Networks}, 9(5), 987--1000.
\newblock \url{https://doi.org/10.1109/72.712178}

\bibitem[Li et~al., 2023]{li2023}
Li, Z., Li, X., Chen, Y., \& Zhang, C. (2023).
\newblock A mechanics-informed machine learning approach for modeling the
  elastoplastic behavior of fiber-reinforced composites.
\newblock {\em Composite Structures}, 323, 117473.
\newblock
  \url{https://doi.org/https://doi.org/10.1016/j.compstruct.2023.117473}

\bibitem[Liang \& Chandrashekhara, 2008]{liang2008}
Liang, G. \& Chandrashekhara, K. (2008).
\newblock Neural network based constitutive model for elastomeric foams.
\newblock {\em Engineering Structures}, 30, 2002--2011.
\newblock \url{https://doi.org/10.1016/j.engstruct.2007.12.021}

\bibitem[Linden et~al., 2023]{linden2023}
Linden, L., Klein, D.~K., Kalina, K.~A., Brummund, J., Weeger, O., \& Kästner,
  M. (2023).
\newblock Neural networks meet hyperelasticity: A guide to enforcing physics.
\newblock {\em Journal of the Mechanics and Physics of Solids}, 179, 105363.
\newblock \url{https://doi.org/https://doi.org/10.1016/j.jmps.2023.105363}

\bibitem[Linka et~al., 2021]{linka2021}
Linka, K., Hillgärtner, M., Abdolazizi, K.~P., Aydin, R.~C., Itskov, M., \&
  Cyron, C.~J. (2021).
\newblock Constitutive artificial neural networks: A fast and general approach
  to predictive data-driven constitutive modeling by deep learning.
\newblock {\em Journal of Computational Physics}, 429, 110010.
\newblock \url{https://doi.org/10.1016/j.jcp.2020.110010}

\bibitem[Linka \& Kuhl, 2023]{linka2023}
Linka, K. \& Kuhl, E. (2023).
\newblock A new family of constitutive artificial neural networks towards
  automated model discovery.
\newblock {\em Computer Methods in Applied Mechanics and Engineering}, 403,
  115731.
\newblock \url{https://doi.org/https://doi.org/10.1016/j.cma.2022.115731}

\bibitem[Liu et~al., 2021]{liu2021}
Liu, Z., Chen, Y., Du, Y., \& Tegmark, M. (2021).
\newblock {\em Physics-augmented learning: A new paradigm beyond
  physics-informed learning}.
\newblock \url{https://arxiv.org/abs/2109.13901}

\bibitem[Marcellini, 1985]{Mar85}
Marcellini, P. (1985).
\newblock Approximation of quasiconvex functions, and lower semicontinuity of
  multiple integrals.
\newblock {\em Manuscripta Math.}, 51(1-3), 1--28.
\newblock \url{https://doi.org/10.1007/BF01168345}

\bibitem[Meyer \& Ekre, 2023]{meyer2023}
Meyer, K.~A. \& Ekre, F. (2023).
\newblock Thermodynamically consistent neural network plasticity modeling and
  discovery of evolution laws.
\newblock {\em Journal of the Mechanics and Physics of Solids}, 180, 105416.
\newblock \url{https://doi.org/https://doi.org/10.1016/j.jmps.2023.105416}

\bibitem[Meyers, 1965]{Mey65}
Meyers, N.~G. (1965).
\newblock Quasi-convexity and lower semi-continuity of multiple variational
  integrals of any order.
\newblock {\em Trans. Amer. Math. Soc.}, 119, 125--149.
\newblock \url{https://doi.org/10.1090/S0002-9947-1965-0188838-3}

\bibitem[Morrey, 1952]{Mor52}
Morrey, Jr., C.~B. (1952).
\newblock Quasi-convexity and the lower semicontinuity of multiple integrals.
\newblock {\em Pacific J. Math.}, 2, 25--53.
\newblock \url{http://projecteuclid.org/euclid.pjm/1103051941}

\bibitem[Morrey, 1966]{Mor66}
Morrey, Jr., C.~B. (1966).
\newblock {\em Multiple integrals in the calculus of variations}, volume Band
  130 of {\em Die Grundlehren der mathematischen Wissenschaften}.
\newblock Springer-Verlag New York, Inc., New York.

\bibitem[Moseley, 2022]{moseley2022}
Moseley, B. (2022).
\newblock Physics-informed machine learning: from concepts to real-world
  applications.
\newblock \url{https://api.semanticscholar.org/CorpusID:254638738}

\bibitem[Neumeier et~al., 2024]{neumeier2024}
Neumeier, T., Peter, M.~A., Peterseim, D., \& Wiedemann, D. (2024).
\newblock Computational polyconvexification of isotropic functions.
\newblock {\em Multiscale Modeling \& Simulation}, 22(4), 1402--1420.
\newblock \url{https://doi.org/10.1137/23M1589773}

\bibitem[Peng et~al., 2021]{peng2021}
Peng, G. C.~Y., Alber, M., Buganza~Tepole, A., Cannon, W.~R., De, S.,
  Dura-Bernal, S., Garikipati, K., Karniadakis, G., Lytton, W.~W., Perdikaris,
  P., Petzold, L., \& Kuhl, E. (2021).
\newblock Multiscale {Modeling} {Meets} {Machine} {Learning}: {What} {Can} {We}
  {Learn}?
\newblock {\em Archives of Computational Methods in Engineering}, 28(3),
  1017--1037.
\newblock \url{https://doi.org/10.1007/s11831-020-09405-5}

\bibitem[Raissi et~al., 2019]{raissi2019}
Raissi, M., Perdikaris, P., \& Karniadakis, G. (2019).
\newblock Physics-informed neural networks: A deep learning framework for
  solving forward and inverse problems involving nonlinear partial differential
  equations.
\newblock {\em Journal of Computational Physics}, 378, 686--707.
\newblock \url{https://doi.org/10.1016/j.jcp.2018.10.045}

\bibitem[Raoult, 1986]{raoult1986}
Raoult, A. (1986).
\newblock Non-polyconvexity of the stored energy function of a saint
  venant-kirchhoff material.
\newblock {\em Aplikace matematiky}, 31(6), 417--419.
\newblock \url{https://doi.org/10.21136/AM.1986.104220}

\bibitem[Settgast et~al., 2020]{settgast2020}
Settgast, C., Hütter, G., Kuna, M., \& Abendroth, M. (2020).
\newblock A hybrid approach to simulate the homogenized irreversible
  elastic–plastic deformations and damage of foams by neural networks.
\newblock {\em International Journal of Plasticity}, 126, 102624.
\newblock \url{https://doi.org/10.1016/j.ijplas.2019.11.003}

\bibitem[Shen et~al., 2004]{shen2004}
Shen, Y., Chandrashekhara, K., Breig, W.~F., \& Oliver, L.~R. (2004).
\newblock Neural network based constitutive model for rubber material.
\newblock {\em Rubber Chemistry and Technology}, 77(2), 257--277.
\newblock \url{https://doi.org/10.5254/1.3547822}

\bibitem[{St. Pierre} et~al., 2023]{pierre2023}
{St. Pierre}, S.~R., Linka, K., \& Kuhl, E. (2023).
\newblock Principal-stretch-based constitutive neural networks autonomously
  discover a subclass of ogden models for human brain tissue.
\newblock {\em Brain Multiphysics}, 4, 100066.
\newblock \url{https://doi.org/https://doi.org/10.1016/j.brain.2023.100066}

\bibitem[Thakolkaran et~al., 2022]{thakolkaran2022}
Thakolkaran, P., Joshi, A., Zheng, Y., Flaschel, M., {De Lorenzis}, L., \&
  Kumar, S. (2022).
\newblock Nn-euclid: Deep-learning hyperelasticity without stress data.
\newblock {\em Journal of the Mechanics and Physics of Solids}, 169, 105076.
\newblock \url{https://doi.org/https://doi.org/10.1016/j.jmps.2022.105076}

\bibitem[Truesdell \& Noll, 1965]{truesdell1965}
Truesdell, C. \& Noll, W. (1965).
\newblock {\em The non-linear field theories of mechanics}.
\newblock Springer .

\bibitem[Wiedemann \& Peter, 2023]{wiedemann2023}
Wiedemann, D. \& Peter, M.~A. (2023).
\newblock Characterization of polyconvex isotropic functions.
\newblock \url{https://arxiv.org/abs/2304.08385}

\bibitem[Zlatic \& Canadija, 2024]{zlatic2024}
Zlatic, M. \& Canadija, M. (2024).
\newblock Recovering mullins damage hyperelastic behaviour with physics
  augmented neural networks.
\newblock {\em Journal of the Mechanics and Physics of Solids}, 193, 105839.
\newblock \url{https://doi.org/https://doi.org/10.1016/j.jmps.2024.105839}

\end{thebibliography}
\end{document}